\documentclass{edp-jp4}

  \usepackage[latin1]{inputenc}
  \usepackage{graphicx}
  \usepackage{amsmath,amssymb}
  \usepackage{ulem}
  \usepackage{indentfirst}
  
  \newcommand{\angstrom}{$\stackrel{\circ}{\rm{A}}$}
  \newcommand{\msol}{{\rm M}_\odot}

\begin{document}



  \title{The Nature of the Low-Metallicity ISM in the Dwarf Galaxy NGC~1569}
  \author{Fr\'ed\'eric Galliano}\address{Service d'Astrophysique, CEA/Saclay, 
                                         L'Orme des Merisiers,
                                         91191 Gif sur Yvette, France}
  \author{Suzanne Madden}\sameaddress{1}
  \author{Anthony Jones}\address{Institut d'Astrophysique Spatiale (CNRS), 
                                 Universite de Paris XI, 91405 Orsay, France}
  \author{Christine Wilson}\address{Department of Physics and Astronomy,
                                    McMaster University, Hamilton, ON L8S 4M1,
                                    Canada}
  \author{Jean-Philippe Bernard}\sameaddress{2}
  \author{Francine Le Peintre}\sameaddress{2}

  \maketitle

\begin{abstract}
We are modeling the spectra of dwarf galaxies from infrared to 
submillimeter wavelengths to understand the nature of the various dust 
components in low-metallicity environments, which may be comparable to the ISM 
of galaxies in their early evolutionary state.
The overall nature of the dust in these environments appears to differ from
those of higher metallicity starbursting systems.
Here, we present a study of one of our sample of dwarf galaxies, NGC~1569,  
which is a nearby, well-studied starbursting dwarf.
Using ISOCAM, IRAS, ISOPHOT and SCUBA data with the D\'esert 
et al (1990) model, we find consistency with little contribution
from PAHs and Very Small Grains and a relative abundance of bigger colder 
grains, which dominate the FIR and submillimeter wavelengths.
We are compelled to use 4 dust components, adding a very cold dust component,
to reproduce the submillimetre excess of our observations. 
\end{abstract}


\section{Introduction}

Dwarf galaxies in our local universe are ideal laboratories for studying the 
interplay between the ISM and star formation in low-metallicity environments
(Hunter \& Gallagher 1989).
They are at relatively early epochs of their chemical evolution, possibly 
resembling distant protogalaxies in their early stages of star formation.
Although only a few metal-poor galaxies have been observed in the MIR using 
ISO, the characteristics of the MIR dust components appear to differ remarkably
from those of normal-metallicity starbursts (Madden 2000).
Whether this is an abundance or composition effect is not yet clear.

For this reason, we are studying their detailed luminosity budget by modeling
their spectral energy distributions (SEDs) from optical to millimeter 
wavelengths, thereby constructing templates to study conditions in primordial
galaxies to help to constrain galaxy evolution models.

The galaxy we present here is NGC~1569.
It is a HI-rich, metal-poor ($Z=0.25\;\rm Z_\odot$) dwarf irregular galaxy 
which lies near the galactic plane at a distance of $(2.2 \pm 0.6)\;\rm Mpc$ 
(Israel 1988).
NGC~1569 is presently in the aftermath of a massive burst of star formation
(Israel 1988, Israel \& De Bruyn 1988, Waller 1991) and exhibits very compact 
HII regions and 2 super-star-clusters (Hunter et al 2000).


\section{Multi-wavelength observations}

  \subsection{Broadband images}

\begin{figure}[htbp]
\begin{center}
  \begin{tabular}{cc}
    \includegraphics[width=8cm]{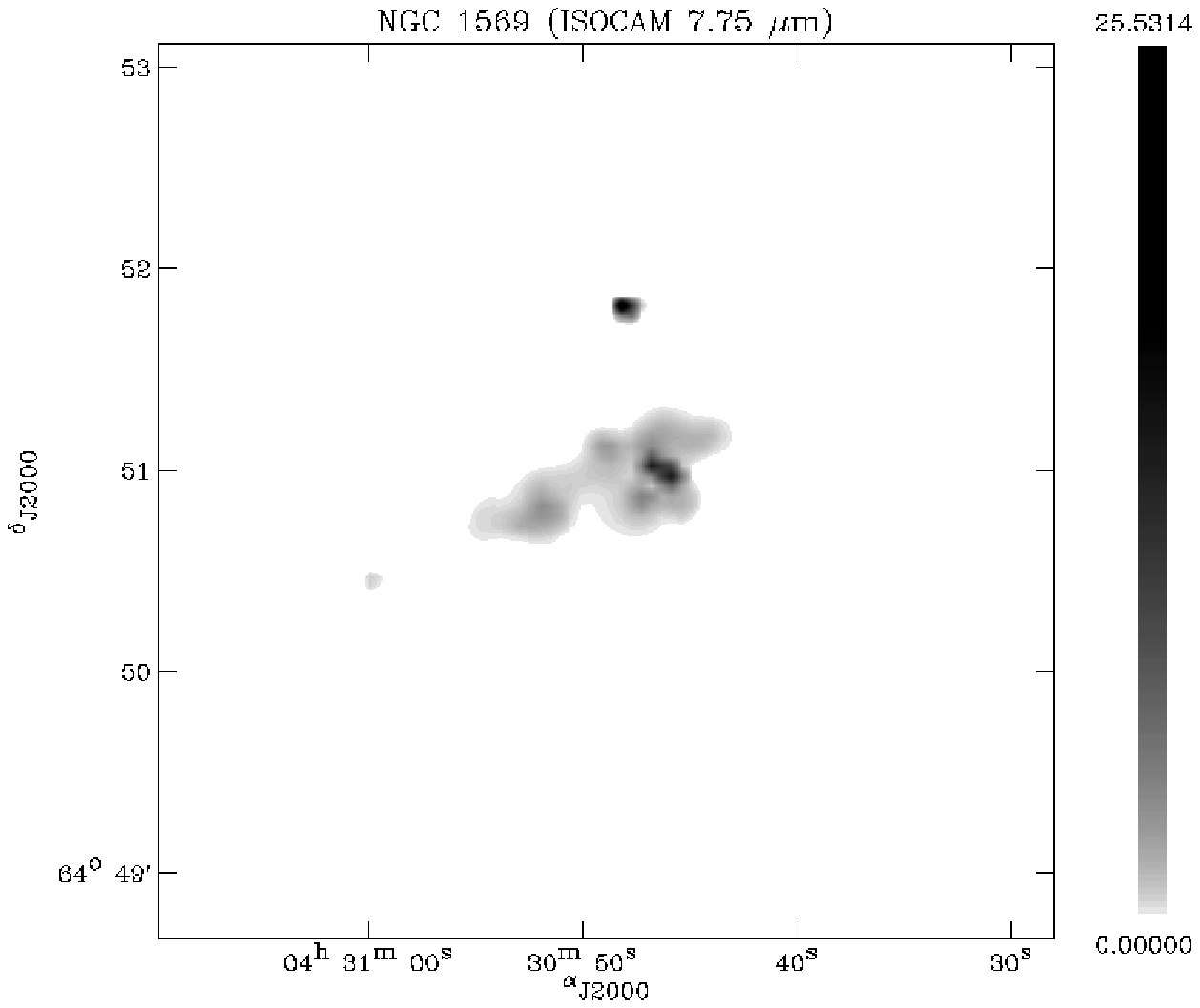} & \hspace{-1.cm}
    \includegraphics[width=8cm]{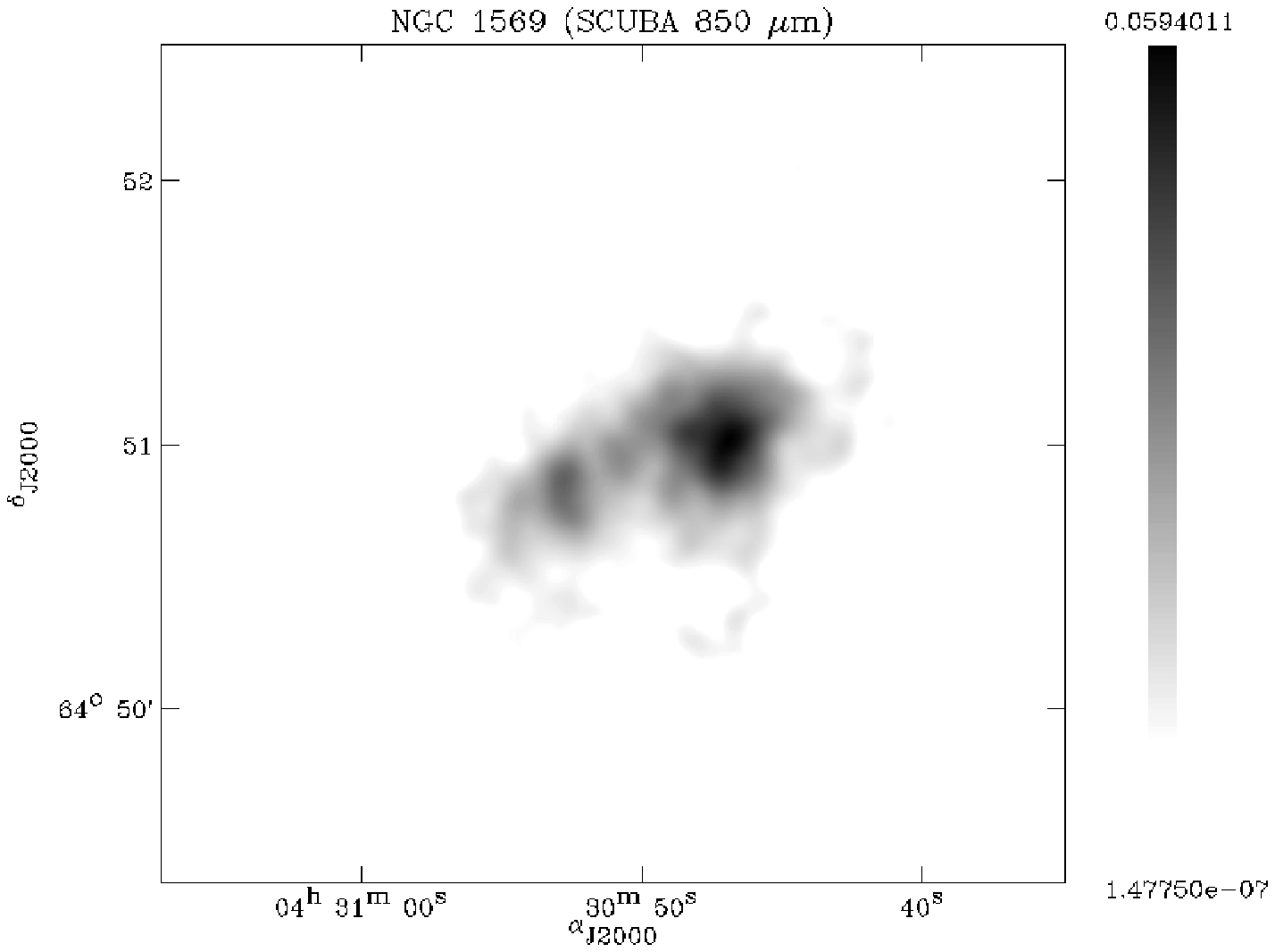} \\
  \end{tabular}
    \caption{\itshape {\bfseries ISOCAM and SCUBA broad-band images.} 
    The image on the left is ISOCAM LW6 (7.75~$\mu m$) calibrated in mJy/pixel
    and has been deconvolved. 
    The image on the right is SCUBA 850~$\mu m$ calibrated in Jy/beam,
    the beam size being $\sim 15''$.}
  \label{fig:obs}
\end{center}
\end{figure}

Figure \ref{fig:obs} shows some of our broadband images of NGC~1569 with 
ISOCAM and SCUBA.
The ISOCAM~7.75~$\mu m$ band traces the aromatic bands (PAH) and
the SCUBA band traces the cold grain continuum.

  \subsection{The MIR spectrum}

\begin{figure}[htbp]
  \begin{center}
    \includegraphics[width=9cm]{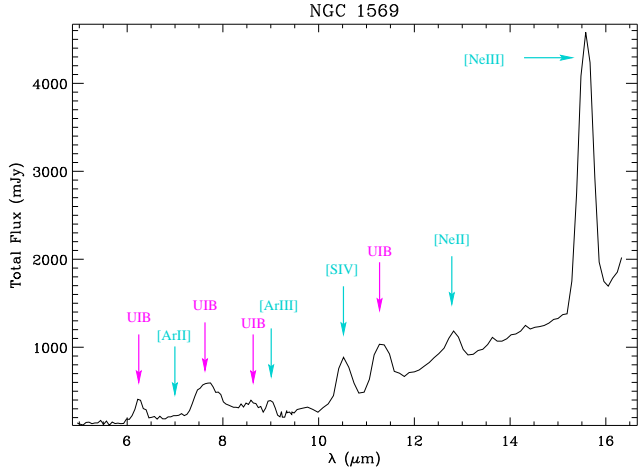}
    \caption{\itshape {\bfseries ISOCAM CVF spectrum.}
             This spectrum is integrated over the whole galaxy.
             The arrows are pointing the ionic lines and the Unidentified 
             Infrared Bands which are the PAHs.}
    \label{fig:cvf}
  \end{center}
\end{figure}

Figure \ref{fig:cvf} shows the ISOCAM CVF spectrum for NGC~1569.
Compared to a normal disk galaxy, there are very strong ionic lines due to 
O-B stars.
The ratio $\rm{[NeIII]/[NeII]} \sim 10$ limits the age of the stellar 
clusters to $\lesssim$~5~Myr.
There are weak aromatic bands, probably due to the destruction of the carriers 
as a consequence of the hard penetrating radiation field in the 
low-metallicity environement.
Moreover, there's a very steeply rising continuum due to very small 
grains emission (see the the D\'esert et al (1990) model) which are 
stochastically heated.


\section{Dust emissivity modeling}

  \subsection{The D\'esert et al (1990) Model}

The D\'esert et al (1990) model empirically computes the dust emissivity 
from near-infrared to submillimetre wavelengths. 
It is a coherent interpretation of both the interstellar extinction
and the infrared emission.
The emission originates from three different components:
  the PAH (Polycyclic Aromatic Hydrocarbon) which are
  aromatic 2-dimensional molecules producing the FUV non-linear part of 
  the extinction curve, 
  the VSG (Very Small Grains) which are 3-dimensional
  carbon grains apparently responsible for the absorption bump at 217.5 nm,
  the BG (Big Grains) which are 3-dimensional silicate grains explaining the 
  NIR and visible rise of the extinction curve.
Each component is described by four parameters: a minimum and 
maximum grain size ($a_-$ and $a_+$), 
a power-law index, $\alpha$, for the size distribution 
and a scaling factor which is the mass abundance relative to 
Hydrogen ($Y = m / m_{\rm{H}}$).
The number density of grains of radius between $a$ and $a + da$ is 
$n (a) \propto a^{-\alpha}$.

  \subsection{Modeling the dust in NGC~1569}

We model the SED of this galaxy, fitting the D\'esert et al (1990) dust model 
to our observations (broadbands and spectrum).
We make the link between the stellar radiation field and the full dust spectrum
by modeling the stellar populations with PEGASE (Fioc, Rocca-Volmerange, 1997)
using UV to NIR data from the literature and MIR ionic 
lines as a further constraint. 
We then use our modeled stellar radiation field as input to the dust model.

\begin{figure}[htbp]
  \begin{center}
    \includegraphics[width=13cm]{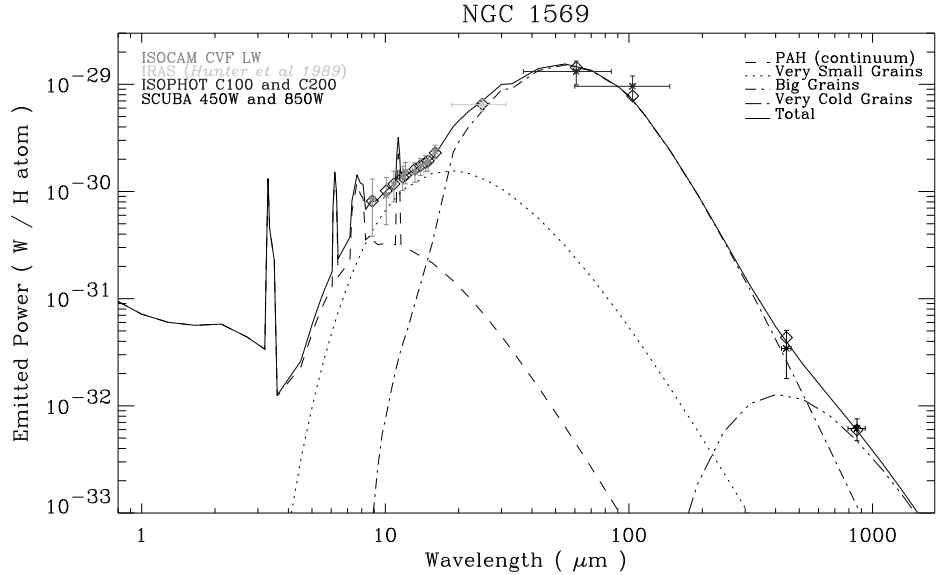}
    \caption{\itshape {\bfseries NGC~1569 Spectral Energy Distribution.}
             The error bars are the observations for the whole galaxy and the 
             lines are the model and its different components.
             Individual PAH bands are not considered in modeling the SED 
             continuum.}
    \label{fig:sed}
  \end{center}
\end{figure}

To obtain the best $\chi^2$ fit, we vary different parameters of the model:
the mass abundances ($Y$), the minimum and maximum sizes ($a_-$ and
$a_+$, except $a_-^{\rm PAH}$), assuming the continuity of the size 
distribution, and the extinction of the synthesized radiation field.
However, if we try to fit our data with only the three components, there 
remains a submillimetre excess we can't explain.
We decide to add a fourth component corresponding probably to Very Cold Grains
(VCGs), modeled by a black body with an emissivity index $\beta$ to reproduce 
this excess and to obtain a correct fit varying the mass abundance of this 
component.
Finally we get $\chi^2=1.3$ for 15 data points and 8 free parameters. 
All the parameters used are summarized in the following table and compared to 
the values for the Milky-Way.
The best fit we obtained is shown on figure \ref{fig:sed}.
\begin{center}
  \begin{tabular}{||l||*{3}{c|c||}l|c||}
    \hline
      &
      \multicolumn{2}{c||}{\bfseries PAH} &
      \multicolumn{2}{c||}{\bfseries VSG} &
      \multicolumn{2}{c||}{\bfseries BG}  & 
      \multicolumn{1}{c}{} & \multicolumn{1}{c||}{\bfseries VCG} \\
    \cline{2-9}
                      &
      Milky           & NGC             & 
      Milky           & NGC             & 
      Milky           & NGC             & 
                      & NGC             \\
                      &
      Way             & 1569            & 
      Way             & 1569            & 
      Way             & 1569            & 
                      & 1569            \\
    \hline
      $Y$             &
      $4.3\, 10^{-4}$ & $2.1\, 10^{-8}$ & 
      $4.7\, 10^{-4}$ & $4.4\, 10^{-6}$ & 
      $6.4\, 10^{-3}$ & $1.9\, 10^{-3}$ & 
      $Y$             & $2.1\, 10^{-3}$ \\ 
    \hline
      $a_-$           &
      4 \angstrom     & 4 \angstrom     & 
      12 \angstrom    & 37 \angstrom    & 
      150 \angstrom   & 47 \angstrom    & 
      T               & 7 K             \\
    \hline
      $a_+$           &
      12 \angstrom    & 37 \angstrom    & 
      150 \angstrom   & 47 \angstrom    & 
      1100 \angstrom  & 8500 \angstrom  & 
      $\beta$         & 1.0             \\
    \hline
      $\alpha$        &
      3               & 3               &
      2.6             & 2.6             &
      2.9             & 2.9             &
      \multicolumn{2}{c}{} \\ 
    \cline{1-7}
  \end{tabular}
\end{center}


\section{Conclusion}

The dust masses deduced from our modeling are 
$m_{\rm PAH}\sim 3 \msol$, 
$m_{\rm VSG}\sim 600 \msol$,
$m_{\rm BG}\sim 2.5\, 10^5 \msol$,
$m_{\rm VCG}\sim 2.7\, 10^5 \msol$ 
and the gas-to-dust mass ratio is $\sim 400$. 
The dust mass deduce from the extinction is $\sim 2.2\, 10^5 \msol$ which is 
the same order of magnitude as the mass deduced from the model.
Our results seem to show that the dust mass in NGC~1569 is mainly concentrated
in cold dust: big grains and very cold grains which both have roughly the same 
mass abundance.
The PAHs as well as the VSGs are very sparse, while the SED is dominated by
bigger grains.



\end{document}